\documentclass[twocolumn,aps,floatfix]{revtex4}
\usepackage{amsmath,amssymb,booktabs,graphicx,verbatim}
\usepackage[colorlinks=true,citecolor=blue,filecolor=blue,linkcolor=blue,urlcolor=blue,pdftex]{hyperref}
\newcommand{\1}[1]{\, \mathrm{#1}} 
\newcommand{\n}[1]{\mathrm{#1}}    
\newcommand{\dd}{\, \mathrm{d}}
\newcommand{\ee}{\mathrm{e}}
\newcommand{\percent}{\%}
\newcommand{\infinity}{\infty}

\newcommand{\arxiv}[1]{\href{http://arxiv.org/abs/#1}{\texttt{arXiv:#1}}}
\begin{document}

\title{Optimum Acceptance Regions for Direct Dark Matter Searches}

\author{Rafael~F.~Lang}
\email{rafael.lang@astro.columbia.edu}
\affiliation{Max-Planck-Institut f\"ur Physik, F\"ohringer Ring 6, D-80805 M\"unchen, Germany, \\ 
now at Columbia Astrophysics Laboratory, Columbia University, New York, NY 10027, USA}

\begin{abstract}
Most experiments that search for direct interactions of WIMP dark matter with a target can distinguish the dominant electron recoil background from the nuclear recoil signal, based on some discrimination parameter. An acceptance region is defined in the parameter space spanned by the recoil energy and this discrimination parameter. In the absence of a clear signal in this region, a limit is calculated on the dark matter scattering cross section. Here, an algorithm is presented that allows to define the acceptance region a priori such that the experiment has the best sensitivity. This is achieved through optimized acceptance regions for each WIMP model and WIMP mass that is to be probed. Using recent data from the CRESST-II experiment as an example, it is shown that resulting limits can be substantially stronger than those from a conventional acceptance region. In an experiment with a segmented target, the algorithm developed here can yield different acceptance regions for the individual subdetectors. Hence, it is shown how to combine the data consistently within the usual Maximum Gap or Optimum Interval framework.
\end{abstract}


\maketitle

\section{Introduction}

Although the existence of dark matter is now well established (see e.g.~\cite{amsler2008b}), we have not yet succeeded in determining its nature. A highly motivated class of models predicts dark matter to be in the form of Weakly Interacting Massive Particles (WIMPs) (see e.g.~\cite{jungman1996}). Direct searches for scatterings of such WIMP dark matter particles off nuclei (see e.g.~\cite{gaitskell2004}) probe and constrain models in highly relevant regions of parameter space. Limits on the cross section of the scattering process are calculated based on the exposure of the experiment as well as events that are observed in an acceptance region. 

The question arises how this acceptance region should be defined. Here, an algorithm is presented that allows to define an optimum acceptance region such that the experiment has the best sensitivity for a given WIMP signal. Hence, one can expect to obtain the most stringent limit on the dark matter scattering cross section. The algorithm is illustrated using recent data from CRESST-II~\cite{angloher2009}, but can be employed in any experiment. It is also shown how to consistently combine data from individual detectors or experiments with differing spectral acceptance in the framework of Yellin's Maximum Gap or Optimum Interval methods~\cite{yellin2002}.

\section{The Optimum Acceptance Region}\label{sec:acceptance}

\begin{figure}[!htbp]
\begin{center}\includegraphics[angle=90,width=1\columnwidth]{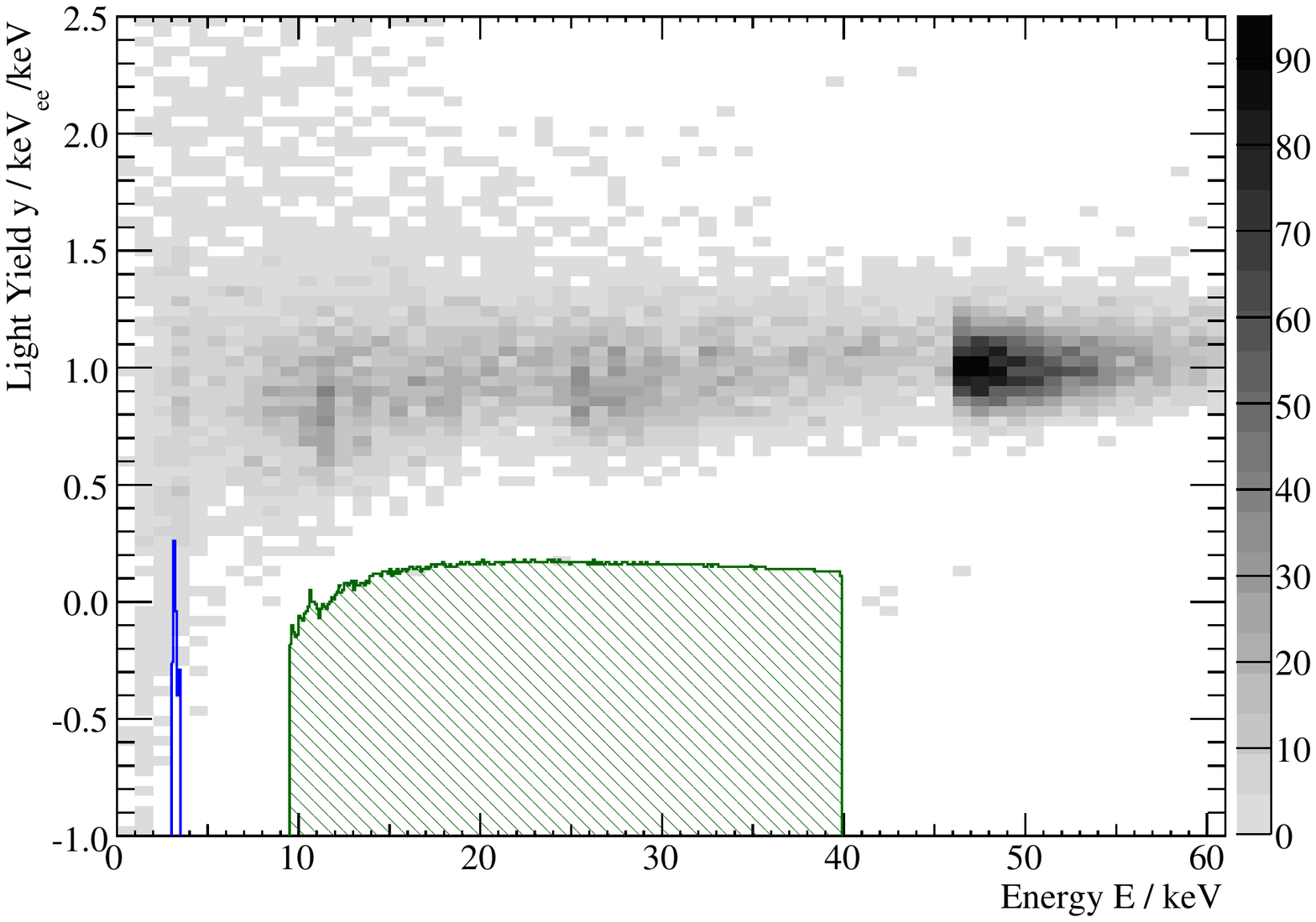}\end{center}
\vspace{-5mm}\caption{Two-dimensional histogram of events collected by one CRESST-II detector module in the light yield--energy-plane $y(E)$ (detector \textsc{Verena} after an exposure of $24.11\1{kg\,d}$~\cite{angloher2009}). Counts per bin are color coded according to the grey scale on the right. The band around a light yield of unity comes from the dominating electron and gamma background. The dark matter induced nuclear recoil signal is expected around zero light yield. Entries with negative light yield originate in the amplitude fitting procedure that allows for negative amplitudes in order to treat noise in an unbiased way~\cite{angloher2005}. The hatched area (green) is an example of the optimum acceptance region for a $100\1{GeV/c^2}$ WIMP, and the area at lower energies (blue) an example for the case of a $10\1{GeV/c^2}$ WIMP.}
\label{fig:VerenaRun30d}\end{figure}

Figure~\ref{fig:VerenaRun30d} shows data collected in the CRESST-II experiment~\cite{angloher2009} that will be used as an example to explain the method. For each particle interaction in the target, the experiment collects two signals: The recoil energy $E$ is taken from the phonon detector, and the light output $L$ is measured by a separate light detector in units of keV electron equivalent $\n{keV_{ee}}$, defined such that $122\1{keV}$ gammas from a $\n{{}^{57}Co}$ calibration deposit $122\1{keV_{ee}}$ in the light detector~\cite{angloher2005}. The light yield $y$ is defined as the ratio of these two parameters, $y\equiv L/E$. It serves as discrimination parameter to distinguish the dominant electron recoil background in the band around $y\approx1\1{keV_{ee}/keV}$ (figure~\ref{fig:VerenaRun30d}) from the nuclear recoil signal which is expected around zero light yield. Other experiments have different discrimination parameters such as the charge yield $Q/E$~\cite{ahmed2009,sanglard2005} or the ratio of delayed and prompt scintillation $S_2/S_1$~\cite{angle2008,alner2007}. The algorithm presented in the following is applicable irrespectively of the nature of the discrimination parameter.

Of course any algorithm that aims to define an acceptance region needs to be blind to the particular distribution of individual events in the data. To facilitate this, the data can be regarded as a background density $\varrho_{\n{b}}(y,E)$ which is calculated based on a parametrization of the data. Hence, $\varrho_{\n{b}}(y,E)$ is a real valued function defined for the whole parameter space $(y,E)$ and is independent of individual signal candidate events. 

In the CRESST-II example considered here, the background can be modeled as a Gaussian band, prominent in figure~\ref{fig:VerenaRun30d}. The energy dependence of the background is taken from the measured spectrum~\cite{lang2009b}. To find the mean $L(E)$ and width $\sigma(E)$ of the background band, the data is fitted with a Gaussian function that allows for the observed scintillator non-proportionality in the mean~\cite{lang2009c}, and with a parametrization for the width $\sigma^2(E) = \sigma_0^2+\sigma_1^2E+\sigma_2^2E^2$ as used by the collaboration~\cite{angloher2005}. 

In addition to $\varrho_{\n{b}}$, there is an expected signal density $\sigma \varrho_{\n{s}}(y,E)$ that is proportional to the WIMP-nucleon scattering cross section $\sigma$. This density is calculated from the position and width of the nuclear recoil band~\cite{angloher2009} and the expected WIMP spectrum $\n{d}\Gamma/\n{d}E$~\cite{donato1998,lewin1996}.

Only events that are found within the acceptance region are considered to calculate a limit. In the CRESST-II example, the acceptance region is parametrized by its upper boundary $y_{\n{max}}(E)$, since the negative boundary can be taken to be at $y_{\n{min}}(E)=\n{const}=-1\1{keV_{ee}/keV}$ which is equivalent to $y_{\n{min}}(E)=-\infinity$ for all practical purposes. For a particular realization of the acceptance region $\{y_{\n{max}}(E), y_{\n{min}}(E)\}$, the expectation values for the observed number of signal and background events are then simple integrals
\begin{eqnarray}
   \langle n_{\n{s}} \rangle & = & \int_0^{\infinity} \dd E \; 
                                   \int_{y_{\n{min}}(E)}^{y_{\n{max}}(E)} \dd y \; \; 
                                   \sigma \varrho_{\n{s}}(y,E) \label{eq:ns}\\
   \langle n_{\n{b}} \rangle & = & \int_0^{\infinity} \dd E \; 
                                   \int_{y_{\n{min}}(E)}^{y_{\n{max}}(E)} \dd y \; \; 
                                   \varrho_{\n{b}}(y,E). \label{eq:nb}
\end{eqnarray}

Given the acceptance region $\{y_{\n{max}}(E), y_{\n{min}}(E)\}$, a particular experiment will observe (after unblinding) a certain number of accepted events $n_{\n{obs}} \in \mathbb{N}$. This number then allows to calculate some upper limit $n_{\n{s,90}}$ still compatible with the data at a stated confidence, typically $90\percent$, often using the Maximum Gap or Optimum Interval methods (section~\ref{sec:combining}). Here it suffices to note that the limit depends on the acceptance region:
\begin{eqnarray}
   n_{\n{s,90}} = n_{\n{s,90}} (n_{\n{obs}}) = n_{\n{s,90}} \big( y_{\n{max}}(E), y_{\n{min}}(E) \big).
\end{eqnarray}
To transfer this number into a limit $\sigma_{90}$ on the cross section, equation~\ref{eq:ns} is evaluated for an arbitrary $\sigma'$, yielding $\langle n_{\n{s}}'\rangle$, and $\sigma_{90}$ can then be calculated according to
\begin{eqnarray}
   \frac{\sigma_{90}}{\sigma'} = \frac{n_{\n{s,90}}}{\langle n_{\n{s}}'\rangle} 
   \qquad \Rightarrow \sigma_{90}     = \frac{\sigma'}{\langle n_{\n{s}}'\rangle } \; n_{\n{s,90}}.
   \label{eq:sigmaconversion}
\end{eqnarray}
Since $\langle n_{\n{s}}'\rangle \propto\sigma'$, the primed variables eventually drop out of the equation again.

The upper limit on the cross section $\sigma_{90}$ will of course depend on the particular realization of the events in a given experiment, so it cannot be used as the objective function. However, the expectation value $\langle\sigma_{90}\rangle$ (the sensitivity of the experiment) is independent of a given experimental outcome and can be used instead. Its value follows from equation~\ref{eq:sigmaconversion}:
\begin{eqnarray}
   \langle \sigma_{90} \rangle & = & \frac{\sigma'}{\langle n_{\n{s}}'\rangle} \; \langle n_{\n{s,90}} \rangle.
\end{eqnarray}

The upper limit $n_{\n{s,90}}$ depends on the number of observed events $n_{\n{obs}}$, which in turn depends on the number of background events $n_{\n{b}}$. The distribution of $n_{\n{b}}$ can be assumed to be Poissonian with expectation value $\langle n_{\n{b}} \rangle$ from equation~\ref{eq:nb}. To calculate the sensitivity $\langle\sigma_{90}\rangle$, we can then simply use the definition of the expectation value:
\begin{eqnarray}
   \langle \sigma_{90} \rangle & = & \frac{\sigma'}{\langle n_{\n{s}}'\rangle} \; 
                                     \sum_{n_{\n{obs}}=0}^{\infinity} n_{\n{s,90}}(n_{\n{obs}})
                                     \; P(n_{\n{obs}}) \nonumber 
                                     \phantom{\ee^{-\langle n_{\n{back}} \rangle}.\label{eq:s90} }\\
                               & = & \frac{\sigma'}{\langle n_{\n{s}}'\rangle} \; 
                                     \sum_{n_{\n{obs}}=0}^{\infinity} n_{\n{s,90}}(n_{\n{obs}})
                                     \; \frac{\langle n_{\n{b}} \rangle^{n_{\n{obs}}}}{n_{\n{obs}}!} \;
                                     \ee^{-\langle n_{\n{b}} \rangle}.\label{eq:s90}
\end{eqnarray}

The sensitivity $\langle\sigma_{90}\rangle$ depends on the acceptance region $\{y_{\n{max}}(E), y_{\n{min}}(E)\}$ through equations~\ref{eq:ns} and~\ref{eq:nb} and is the objective function of choice.

An algorithm for finding the optimum acceptance region varies $y_{\n{max}}(E)$ and $y_{\n{min}}(E)$ until $\langle\sigma_{90}\rangle$ is minimal. To this end, the $(y,E)$ plane is binned and varied on its borders: For each energy bin, the acceptance region is increased or decreased as long as it improves the sensitivity. The algorithm is iterated until the sensitivity converges.

\begin{figure}[!htbp]
\begin{center}\includegraphics[angle=90,width=1\columnwidth,clip,trim=0 0 180 0]{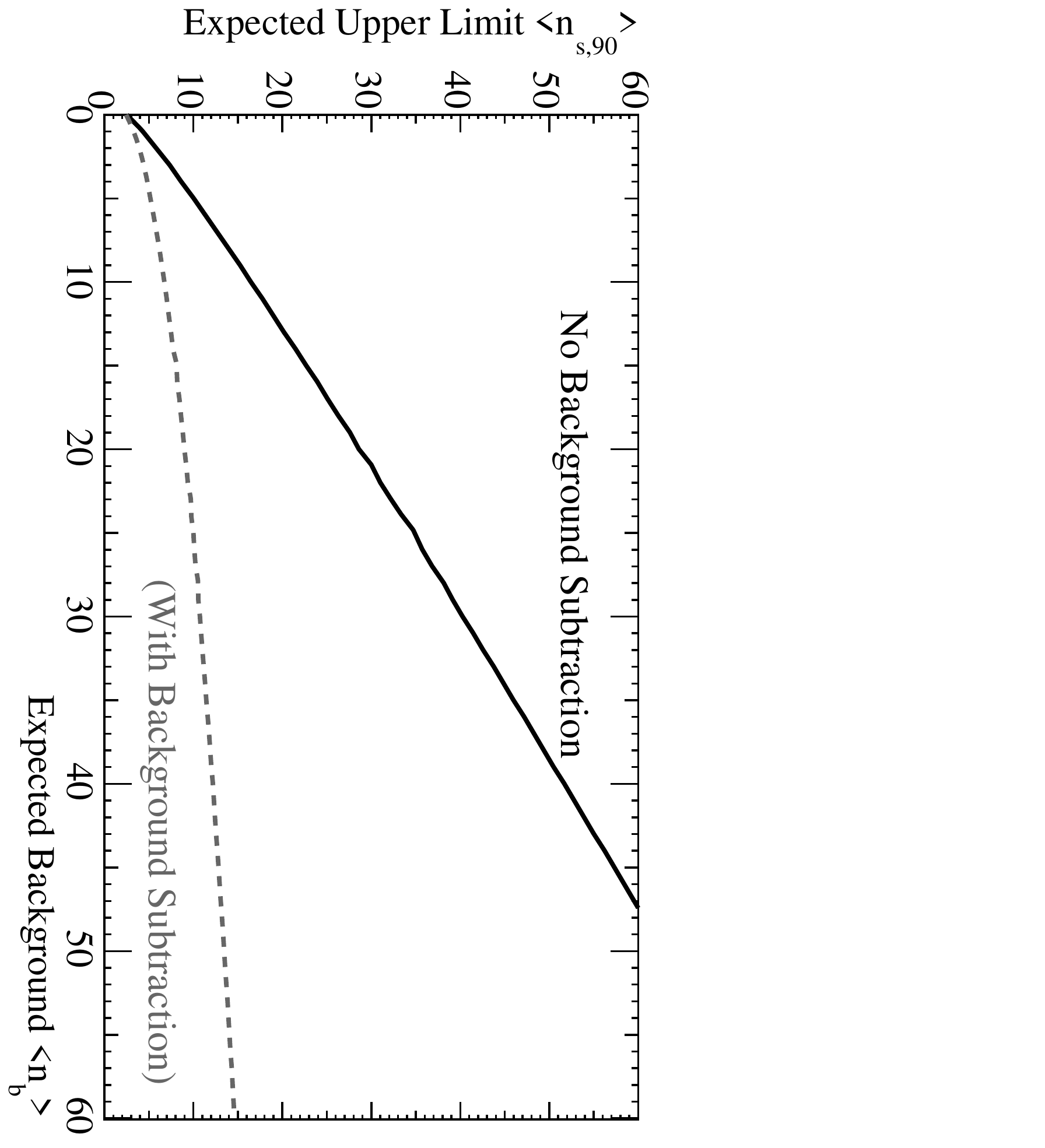}\end{center}
\vspace{-5mm}\caption{The expectation value $\langle n_{\n{s,90}} \rangle$. The solid curve gives the employed curve based on Poisson statistics as shown in equation~\ref{eq:s90}; dashed the (not used) variant in which the background is taken into account using the ordering scheme of Feldman and Cousins.}
\label{fig:n90Graphg}\end{figure}

Some technical remarks will help to implement the algorithm. Although $\ee^{-\langle n_{\n{b}} \rangle}$ can be drawn out of the sum in equation~\ref{eq:s90}, this does not mean that $\langle \sigma_{90} \rangle \rightarrow 0$ as $\langle n_{\n{b}} \rangle\rightarrow\infinity$, since $n_{\n{s,90}}$ counteracts. Instead, the sum is very well behaved, as can be seen in figure~\ref{fig:n90Graphg}, where $n_{\n{s,90}}(n_{\n{obs}})$ is calculated using Poisson statistics. Had one precise knowledge of the background, $n_{\n{s,90}}$ could also be calculated e.g.\ from the Feldman-Cousins scheme~\cite{feldman1998} (dashed line in figure~\ref{fig:n90Graphg}), but this is not used here. Since the sum is well behaved, it can be tabulated for integer $\langle n_{\n{b}} \rangle$ and then interpolated to speed up its computation. Given the simple Gaussian behavior of the signal and background densities $\varrho_{\n{s}}$ and $\varrho_{\n{b}}$, it is not surprising that the objective function is also well behaved, so the algorithm rapidly converges to the optimum acceptance region. 
Eventually, this region can be decreased again by a small amount (e.g. by $0.1\percent$ of the accepted WIMP spectrum) to prevent numerical ambiguities from showing up at higher energies. The optimum acceptance region then naturally extends to a maximum energy above which the acceptance is zero, $y_{\n{max}}=y_{\n{min}}$.

Since the acceptance region is calculated using the signal expectation $\sigma \varrho_{\n{s}}(y,E)$, it will be different for each WIMP model that is probed. In particular, a separate acceptance region is calculated for each WIMP mass. For the CRESST-II example considered here, two optimum acceptance regions are shown in figure~\ref{fig:VerenaRun30d} for generic WIMPs with masses of $10\1{GeV/c^2}$ and $100\1{GeV/c^2}$. For the latter, the acceptance region follows the resolution of the light channel. The acceptance regions are not smoothly bounded but show small dents. This is because the background density $\varrho_{\n{b}}$ needed to evaluate equation~\ref{eq:nb} is calculated using the observed spectrum, smeared in $y$ by the resolution of the light channel. Hence, the acceptance region is reduced for example around $11\1{keV}$ where a gamma line appears in the spectrum~\cite{lang2009b}.

It may be surprising to note the shape of the optimum acceptance region for low mass WIMPs: It is in fact beneficial to \textit{increase} the acceptance toward lower recoil energies. This is due to the recoil spectrum being a sharply falling exponential, confined to the lowest energies. Hence, the increasing electron/gamma background in this parameter region is being overcompensated for.

\begin{figure}[!htbp]
\begin{center}\includegraphics[angle=90,width=1\columnwidth]{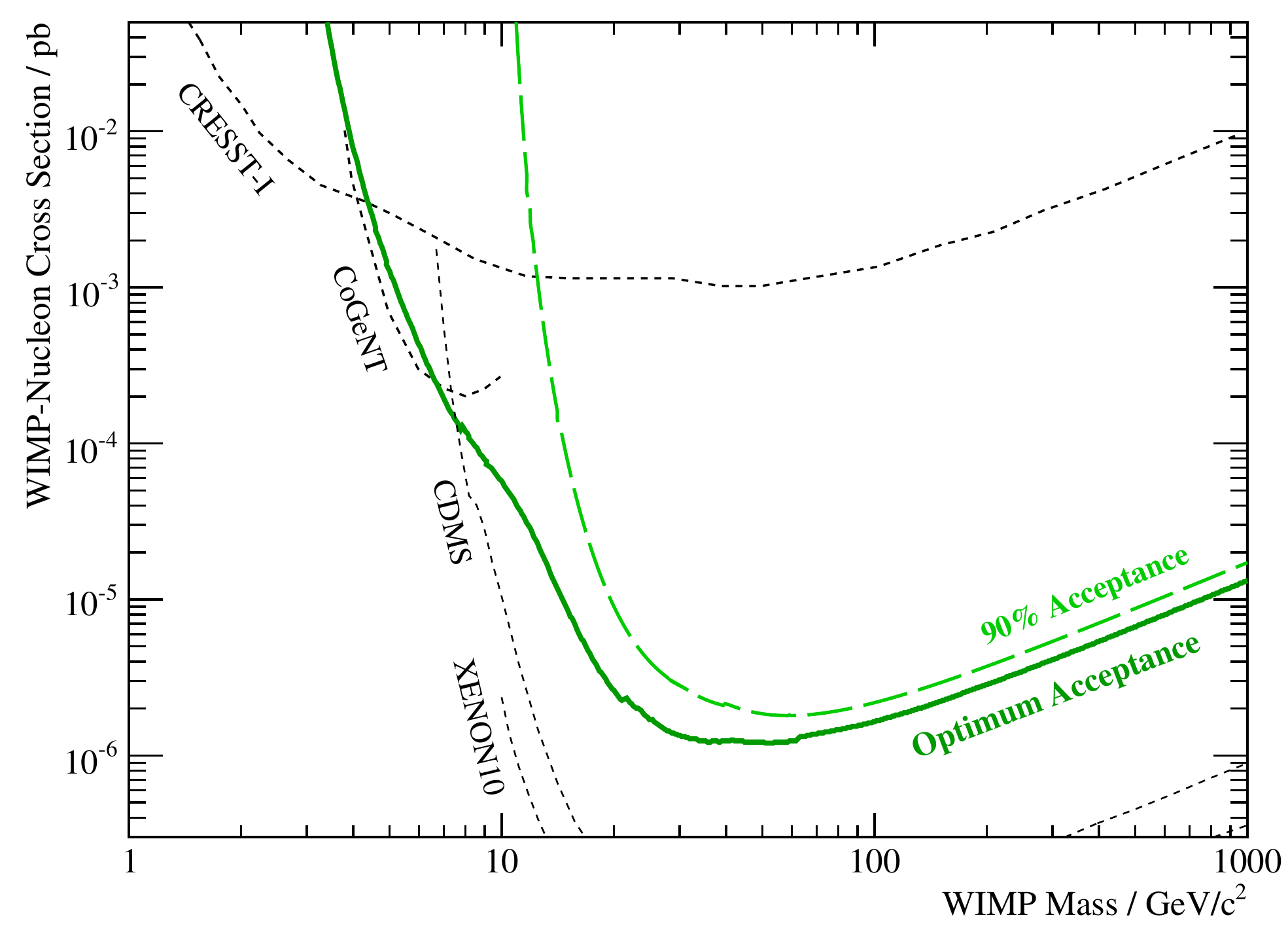}\end{center}
\vspace{-5mm}\caption{Exclusion plot on the WIMP-nucleon scattering cross section as function of WIMP mass, demonstrating the impact of the optimum acceptance region. The thick solid line (green) is the limit obtained with the optimum acceptance regions from the CRESST-II dataset discussed here as an example. The thick dashed line (light green) is the limit from the same dataset obtained by a one-sided $90\percent$ tungsten recoil acceptance region in the energy interval $[10;40]\1{keV}$~\cite{angloher2009}. Limits from some other experiments are also shown: CRESST-I~\cite{angloher2002}, CoGeNT~\cite{aalseth2008}, CDMS~\cite{ahmed2009} and XENON10~\cite{angle2008}.} 
\label{fig:LimitsVerenaRun30h}\end{figure}

The improvement brought by this algorithm can be illustrated calculating a limit based on the above data set. To this end, an analysis threshold of $3\1{keV}$ is imposed to stay well above detection threshold~\cite{lang2009d}, and the same WIMP expectation as used by the CRESST collaboration is probed~\cite{angloher2005}. In particular, the WIMPs are expected to be distributed in the Milky Way in an isothermal halo with a local density of~$0.3\1{GeV/c^2/cm^3}$, through which Earth moves with a velocity of~$230\1{km/s}$~\cite{donato1998}. Recoils with high momentum transfers are suppressed by the form factor, which is taken to be the one introduced by Helm~\cite{helm1956}. Cross sections are normalized to a single nucleon by taking the assumed spin-independent enhancement $\propto A^2$ on the different elements of the target into account. Limits on the cross section of a coherent WIMP-nucleon scattering process are calculated using the optimum acceptance region derived above and Yellin's Optimum Interval method~\cite{yellin2002}. Figure~\ref{fig:LimitsVerenaRun30h} compares the limit from the optimum acceptance region with the limit obtained from this dataset by the CRESST collaboration. There, the acceptance is defined to contain $90\percent$ of the tungsten recoils in the energy interval $[10;40]\1{keV}$~\cite{angloher2009}. For comparison, some limits from other experiments are also shown.

At high WIMP masses, the optimum acceptance region gives a limit that is about $20\percent$ stronger than with the acceptance region previously used by the CRESST collaboration. For low WIMP masses, the optimum acceptance region results in drastically improved limits. For example, for a $10\1{GeV/c^2}$ WIMP the improved limit is more than five orders of magnitude stronger than the one derived from a constant acceptance, that is to say, the optimum acceptance region allows to infer information about WIMPs in otherwise completely inaccessible regions of parameter space.

\section{Combining Data from Different Detectors}\label{sec:combining}

The optimum acceptance region will be distinct for each detector, given its particular background and resolution. Thus, the question arises for segmented-target experiments how to combine the individual subdetectors within the Maximum Gap or Optimum Interval methods. Within these methods, the computed limit depends on integrals over the accepted signal spectrum $\dd \Gamma_{\n{a}}/\dd E$ of the form
\begin{eqnarray}
   n_{\n{i}} := \int_{E_i}^{E_{\n{i+1}}} \frac{\dd \Gamma_{\n{a}}}{\dd E} \dd E
\end{eqnarray}
between two observed events in the acceptance region at energies $E_i$. To facilitate the combination of different detectors, the energy coordinate $E$ is transformed to a new variable $\eta$ according to the integral transform
\begin{eqnarray}
   \eta(E) := \frac{1}{\mathcal{N}} \int_{0}^E \frac{\dd \Gamma_{\n{a}}}{\dd E'} \dd E'\label{eq:transformation}
\end{eqnarray}
which is a bijective transformation that leaves the calculated limit unchanged and occurs naturally within the Maximum Gap or Optimum Interval frameworks. The normalization constant $\mathcal{N}$ is chosen as the accepted-spectrum weighted exposure
\begin{eqnarray}
   \mathcal{N} = \int_{0}^{\infinity} \frac{\dd \Gamma_{\n{a}}}{\dd E'} \dd E'.
\end{eqnarray}
In this new energy variable $\eta$, the accepted signal spectrum is just a constant, which can be most easily seen manipulating differentials: 
\begin{eqnarray}
\frac{\dd \Gamma_{\n{a}}}{\dd \eta}   
                                      =   \frac{\frac{\dd \Gamma_{\n{a}}}{\dd E}} {\frac{\dd \eta}{\dd E}}
                                      =   \frac{\frac{\dd \Gamma_{\n{a}}}{\dd E}} {\frac{\dd}{\dd E} 
                                          \;\frac{1}{\mathcal{N}}\; \int \frac{\dd \Gamma_{\n{a}}}{\dd E}\dd E}
                                      =   \frac{\frac{\dd \Gamma_{\n{a}}}{\dd E}} {\frac{1}{\mathcal{N}}
                                          \;\frac{\dd \Gamma_{\n{a}}}{\dd E}}
                                      = \mathcal{N}. \nonumber
\end{eqnarray}

With this transformation, all possible differences of individual detectors have been mapped into the interval $\eta=[0,1]$ and the single number $\mathcal{N}$. Events of all detectors are distributed within $\eta=[0,1]$, and $\mathcal{N}$ is a measure of the exposure and acceptance of each detector. Therefore, for each expected WIMP spectrum (and in particular for each WIMP mass), individual detectors can now be joined to give the combined limit: The desired Maximum Gap or Optimum Interval method is simply applied to the energy interval $[0,1]$ considering the observed events from all detectors. The summed exposure is obtained by adding the individual $\mathcal{N}$ up.

\section{Conclusions}

Previously, direkt dark matter searches would constrain various WIMP models using one common acceptance region. Here it was shown that by optimizing the acceptance region for each WIMP model, one can improve the sensitivity of an experiment by orders of magnitude. This has been demonstrated with recent data from CRESST-II as an example, where a drastically improved limit resulted in particular for low mass WIMPs. At the same time, the algorithm introduced here removes the ambiguity from defining the acceptance region in a rather ad hoc way. It was shown how to make full use of this optimization within the Maximum Gap or Optimum Interval frameworks to achieve a combined limit for individual subdetectors of a segmented-target experiment, or different experiments altogether.

\section*{Acknowledgements}

I am thankful to my colleagues Jens Schmaler, Dieter Hauff and Franz Pr\"obst for useful discussions.

\end{document}